\documentclass[pra,showpacs,superscriptaddress]{revtex4}
\newcommand{\be}{\begin{equation}}

\newcommand{\ee}{\end{equation}}

\newcommand{\bea}{\begin{eqnarray}}

\newcommand{\eea}{\end{eqnarray}}

\usepackage{graphicx}

\begin{document}

\title{Mean-field theory for Bose-Hubbard Model under a magnetic field}

\author{M. {\"O}. Oktel}

\affiliation{Bilkent University, Department of Physics, 06800
Bilkent, Ankara, Turkey}

\email{oktel@fen.bilkent.edu.tr}

\author{M. Ni{\c t}{\u a}}

\affiliation{Institute of Physics and Technology of Materials,
P.O. Box MG7, Bucharest-Magurele, Romania}

\author{B. Tanatar}

\affiliation{Bilkent University, Department of Physics, 06800
Bilkent, Ankara,  Turkey}

\date{\today}

\begin{abstract}

We consider the superfluid-insulator transition for cold bosons under
an effective magnetic field. We investigate how the applied magnetic
field affects the Mott transition within mean-field theory and find
that the critical hopping strength $(t/U)_c$, increases with
the applied field. The increase in the critical hopping follows the
bandwidth of the Hofstadter butterfly at the given value of
the magnetic field. We also calculate the magnetization and superfluid
density within mean-field theory.

\end{abstract}


\maketitle



\section{Introduction}

One of the most interesting developments in ultra-cold atom
physics is the study of neutral atoms in optical lattices
\cite{ol}.
An optical lattice is prepared by creating a periodic potential
utilizing a standing wave of light, and optical lattices in one,
two and three dimensions have been realized experimentally.

The cooling, trapping and coherent manipulation of the atomic
motion by their interaction with light has been established by
numerous investigations in the field of atom interferometers
\cite{burgbacher}, matter-wave superradiance \cite{inouye571:99},
matter wave parametric amplifiers \cite{kozuma, inouye641:99} and
others \cite{nature}.
 One should
mention also the opportunities garnered by using the ultra-cold
alkali atoms as  quantum computers \cite{jane}, or by using the
Mott insulating state of neutral bosonic atoms for detection of
quantum entanglement \cite{mandel,hensinger,moura}.The custom-made
trapping potentials in the optical lattice has also opened a venue
to study many condensed matter problems, such as the Mott
insulator transition experimentally realized by Greiner {\it et
al}.\cite{greiner}

Although many different regimes exist for optical lattice
experiments, one that is quite interesting from a theoretical
point of view is that of a deep lattice with few particles per
lattice site. If fermions are used instead of bosons, these
experiments may lead to direct realization of many correlated
electron model Hamiltonians such as the Hubbard model of high
temperature superconductivity or lattice Quantum Hall models\cite{qhe}.

In this work, we concentrate on bosons, and assume that at each
lattice site there is only one available state (that is
equivalent to requiring the first excited state at each lattice site
to be sufficiently high in energy). In this case the Hamiltonian
is\cite{toolbox}
\be
\label{bh}
{\cal H} = - t \sum_{<ij>} (a^\dagger_i a_j + a^\dagger_j a_i) +
\frac{U}{2} \sum_{i} (n_i - 1) n_i - \mu \sum_i n_i\, \ee where
$a_i$ is the annihilation operator at site $i$ and
$n_i=a^\dagger_i a_i$ is the number operator at site $i$. The
first term corresponds to hopping between different lattice sites
and for practical purposes only nearest neighbor hopping is
important, so the sum $<ij>$ is carried over the nearest
neighbors. Second term is the particle-particle interaction and
the last term is the chemical potential. This is the widely
studied Bose-Hubbard Hamiltonian \cite{fisher,toolbox,jbc98}.

The strong tunneling limit between optical lattice sites ($U/t
<<1$) corresponds to the superfluid (SF) phase. Changing the laser
intensity with increasing depth of the optical potential the
atomic waves become more localized and the on-site interaction $U$
increases at the same time with reduction of the tunneling
parameter $t$ \cite{jbc98}. The system is driven to a Mott
insulator (MI) phase and loses long range phase coherence. In
general, if the interaction is strong enough the system prefers a
particle number that is commensurate with the number of lattice
sites and the system goes into the insulating phase. The strength
of interaction $U$ needed for this transition is roughly the
bandwidth of the noninteracting system $2 z t$, where $z$ is the
number of nearest neighbors.

A much less studied problem is that of the Bose-Hubbard
Hamiltonian under a magnetic field. Experimentally, of course, the
bosons used in cold gas experiments are uncharged and would not be
directly affected by an external magnetic field. However, recent studies
have shown in detail how a magnetic Hamiltonian , or in general
effective electromagnetic fields can be generated for atoms in
optical lattices using an external time varying electric
field \cite{jzo03}, by
an oscillating quadrupole potential together with a periodic
modulation of the tunnelling between lattice sites \cite{sdl04} or
using more complicated laser configurations \cite{ejm04}. These
investigations suggest that it may be possible to study such
systems with reasonable improvements on already functioning
experiments.  An effective magnetic field can also be created by
rotating the optical lattice, and cancelling the centrifugal force
of the rotation by an external quadratic trap \cite{cornell}.
We also note the
two recent papers which study how the artificial external
non-Abelian gauge potentials can be created for cold atom systems
\cite{osterloh05,ruseckas05}.

In this work, we  assume
that we have a two dimensional square lattice in the $x-y$ plane,
under a magnetic field in the $z$ direction. We also consider that the
"charged bosons" are interacting only
when they are on the same lattice site and the temperature of the
system is set to zero. In this case the Hamiltonian is

\bea\label{bhh}
{\cal H}= -t \sum_{n,m} \left[ a^\dagger_{nm} a_{(n+1)m}
+ e^{i 2 \pi \varphi n} a^\dagger_{nm}
a_{n(m+1)} + h.c. \right] \\
+ \frac{U}{2} \sum_{n,m} (a^\dagger_{nm} a_{nm} -1)
a^\dagger_{nm} a_{nm}
- \mu \sum_{n,m} a^\dagger_{nm} a_{nm}\, .
\eea

Here, we label every site of the lattice $i,j$ by two integers
$i=(n_i,m_i)=(n,m)$, one integer ($n$) along the
$x$ axis, the other ($m$) along the $y$ axis, and choose the
gauge to be $\vec{A}= x \hat{y}$. The first term is the usual
hopping term, where hopping along the $y$ axis gets a
phase shift due to the presence of the magnetic field. Magnetic
field affects the system through the parameter $\varphi$ with
\be
\varphi = B l^2/\varphi_0\,
\ee
where $l$ is the lattice spacing and $\varphi_0$ is the flux quantum.
Thus, the parameter $\varphi$ measures the magnetic flux
per unit cell of the lattice in units of flux quantum . Second and
third terms are interactions and chemical potential, respectively.
Even at $U=0$, the noninteracting limit of this Hamiltonian shows
interesting results; the energy spectrum at $U=0$ is known
as the Hofstadter butterfly \cite{butter}. Most important aspect
of the non-interacting problem is that the bandwidth depends
critically on $\varphi$, and gaps open up or close in a self-similar
manner. With such a complicated single particle spectrum,
it is not at all clear how the presence of the magnetic field will
change the Mott transition, or whether more exotic phases can be found.

We believe that with the possibility of experimental realization,
it is of importance to study this model more closely and
understand its rich phase diagram. In this work, we concentrate on
the superfluid-insulator transition and investigate the effect of
the external magnetic field on the phase boundary. Our mean-field
approach is not capable of capturing possible correlated phases,
however it may serve as a basis for more detailed investigation of
the model.

We find that the Mott insulating phases become more stable under
the applied magnetic field, an expected effect as one of the most
important effects of the magnetic field would be to localize
particles further. More importantly, we find that the critical
hopping to interaction ratio $t/U$, roughly follows the bandwidth
of the Hofstadter butterfly. We also calculate the magnetization and
the superfluid density within mean-field theory.

In the rest of this paper we first outline our calculational scheme
of solving the Bose-Hubbard model under a magnetic field within the
mean-field approach. We then present our results on the phase
diagram identifying the superfluid and insulating regions.
We conclude with a brief summary of our main results.

\section{Mean-Field Approach}

Our calculations are based on the mean-field approach of the
Bose-Hubbard Hamiltonian \cite{sheshadri93},
by considering the following
decoupling formula for the product of the two Bose field operators:

\bea\label{decoupling}
a^\dagger_{nm} a_{(n+1)m}=\langle a^\dagger_{nm}\rangle a_{(n+1)m} +
                 a^\dagger_{nm}\langle a_{(n+1)m}\rangle-
                 \langle a^\dagger_{nm}\rangle\langle a_{(n+1)m}\rangle\, .
\eea
The average value $\langle a^\dagger_{nm}\rangle$ represents
the order parameter $\Psi_{nm}$ that accounts for the
insulator-superfluid transition.
It is equal to zero on the insulator side of the transition when
the ground state of the system
has a definite particle number on every site of the lattice,
and has a nonzero value for the superfluid state when there
are large  quantum fluctuations of the atom number in the optical
lattice. In this case $|\Psi_{nm}|^2$ represents the local density
of the atoms in the condensate state.

Using Eq.\,(\ref{decoupling}) the Bose-Hubbard
Hamiltonian given in Eq.\,(\ref{bhh}) turns into a sum of the following
single-site terms:
\bea\label{mf}
{\cal H}_{nm}^{MF}=&&-t \left[ \Psi_{(n+1)m}^{*} a^\dagger_{nm} + \Psi_{(n-1)m}^{*} a^\dagger_{nm}
                    +e^{ i 2 \pi \varphi n} \Psi_{n(m+1)}^{*} a^\dagger_{nm}
          +e^{-i 2 \pi \varphi n} \Psi_{n(m-1)}^{*} a^\dagger_{nm}+ \hbox{h.c.} \right] \nonumber \\
          &&+ U (n_{nm} -1) n_{nm}  - \mu  n_{nm} +C_{nm}\, ,
\eea
where $n_{nm}$ is the single-site density operator $a^\dagger_{nm}a_{nm}$
and $C_{nm}$ is a constant energy term.

The matrix elements of the mean-field Hamiltonian ${\cal H}_{nm}^{MF}$
in the occupation number base
$|N_{nm}\rangle$ are given by:
\bea\label{med}
\langle N_{nm}|{\cal H}_{nm}^{MF}|N_{nm}\rangle=
                      &&\frac{1}{2}UN_{nm}(N_{nm}-1)-\mu N_{nm} +C_{nm},\\\label{men}
\langle N_{nm}+1|{\cal H}_{nm}^{MF}|N_{nm}\rangle=
   &&-t\sqrt{N_{nm}+1}\left[\Psi_{(n+1)m}^{*}+\Psi_{(n-1)m}^{*}
                         +e^{ i 2 \pi \varphi n} \Psi_{n(m+1)}^{*}+
                         e^{-i 2 \pi \varphi n} \Psi_{n(m-1)}^{*}\right]\, ,
\eea
where we used the property of the Bose field operators
$c|N+1\rangle=\sqrt{N+1}|N\rangle$
and $c^\dagger|N\rangle=\sqrt{N+1}|N+1\rangle$. All other matrix
elements are zero, except the conjugate elements of Eq.\,(\ref{men}).
We note that the occupation number $N_{nm}$  above
varies from $0$ to $\infty$ and they are referred to the site
$(nm)$ of the optical lattice. We diagonalize the Hamiltonian
Eq.(\ref{mf}) in a truncated
basis $|N_{nm}\rangle$ with $N_{nm}=0... N_{max}$ and calculate the
ground state of the mean-field Hamiltonian:
\be\label{ssgs}
|G^{nm}\rangle=\sum_{N=0}^{N_{max}}\alpha_N^{nm}|N_{nm}\rangle\, ,
\ee
with the coefficients $\alpha_N^{nm}$ corresponding to the lowest
eigenvalue of the matrix of Eq.\,(\ref{mf}) in the truncated base.

The order parameter corresponding to the ground state given by
Eq.\,(\ref{ssgs}) will be
\bea\label{mfop}
\Psi_{nm}=\langle G^{nm}|a_{nm}^\dagger|G^{nm}\rangle=
\sum_{N=0}^{N_{max}-1}\alpha_N^{nm*}\alpha_{N+1}^{nm}\sqrt{N+1}\, .
\eea

For a given truncated basis the equations of the finite Hermitian
matrix in Eq.\,(\ref{med}) and Eq.\,(\ref{men}) and the formula for
the SF order parameter Eq.\,(\ref{mfop}) represent a set of
self-consistent equations that
give the solution of the ground state of the single site
Hamiltonian Eq.\,(\ref{mf}) and the order parameters $\Psi_{nm}$
in the mean-field approximation.

The numerical
calculations are repeated with increasing values of the dimension
$N_{max}$ of the truncated basis to attain convergence of the solution.
In the mean-field approximation, the ground state of the Bose-Hubbard
Hamiltonian Eq.\,(\ref{bhh}) is given by the direct product of the
single site ground states of Eq.\,(\ref{ssgs}):

\bea\label{gs}
|G\rangle=\prod_{nm}|G^{nm}\rangle.
\eea

For a given ground state Eq.\,(\ref{ssgs}) the probability for
the single site operator $n_{nm}$ to take the value $N$ will be
given by the square of the corresponding developing coefficient
$\alpha_N^{nm}$. The average single site occupation number
denoted with $\rho(nm)$ is equal to:
\bea\label{mfn}
\rho(nm)=\langle n_{nm}\rangle=
\sum_{N=1}^{N_{max}}|\alpha_N^{nm}|^2 N\, ,
\eea
and the condensate component of the superfluid density, within
mean-field theory,  on the site $nm$ is:
\bea\label{mfns}
\rho_s(nm)=|\Psi_{nm}|^2\, .
\eea

\begin{figure}
\includegraphics[scale=0.7]{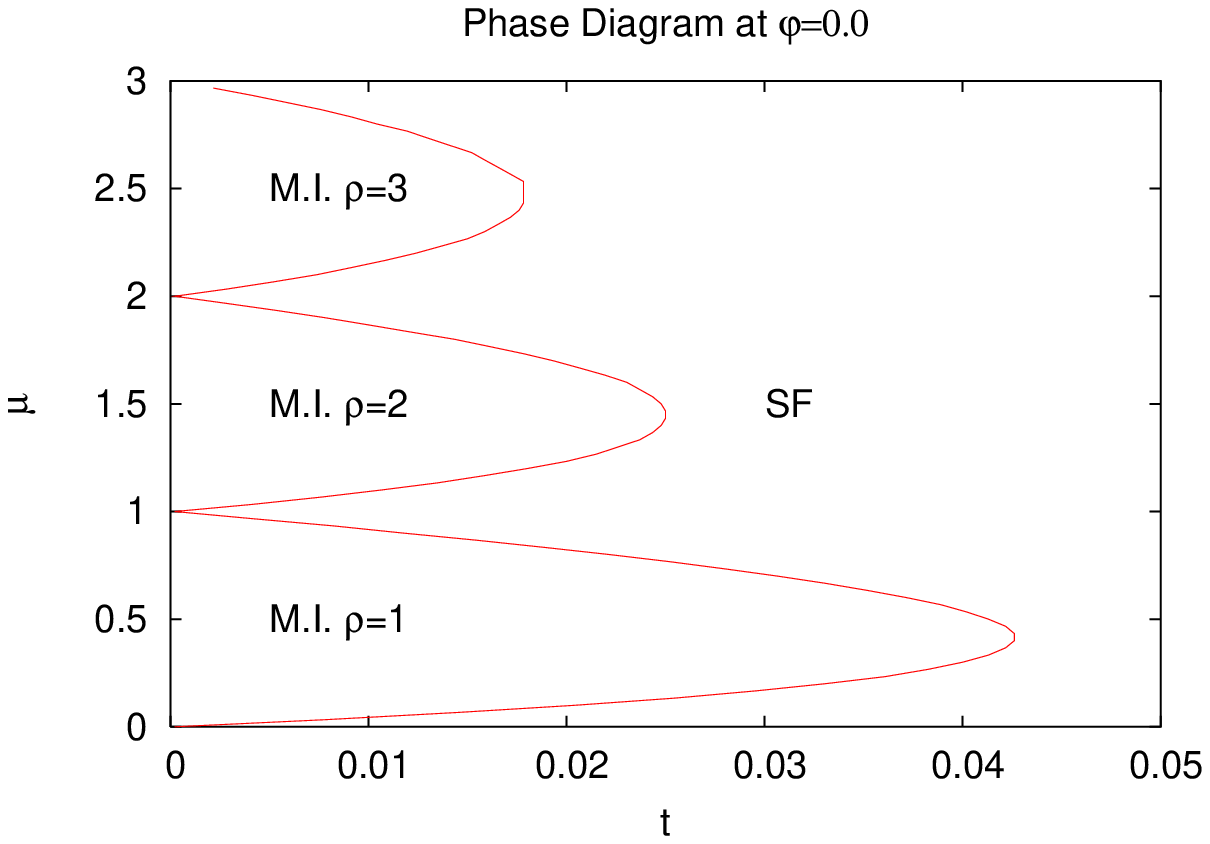}
\includegraphics[scale=0.7]{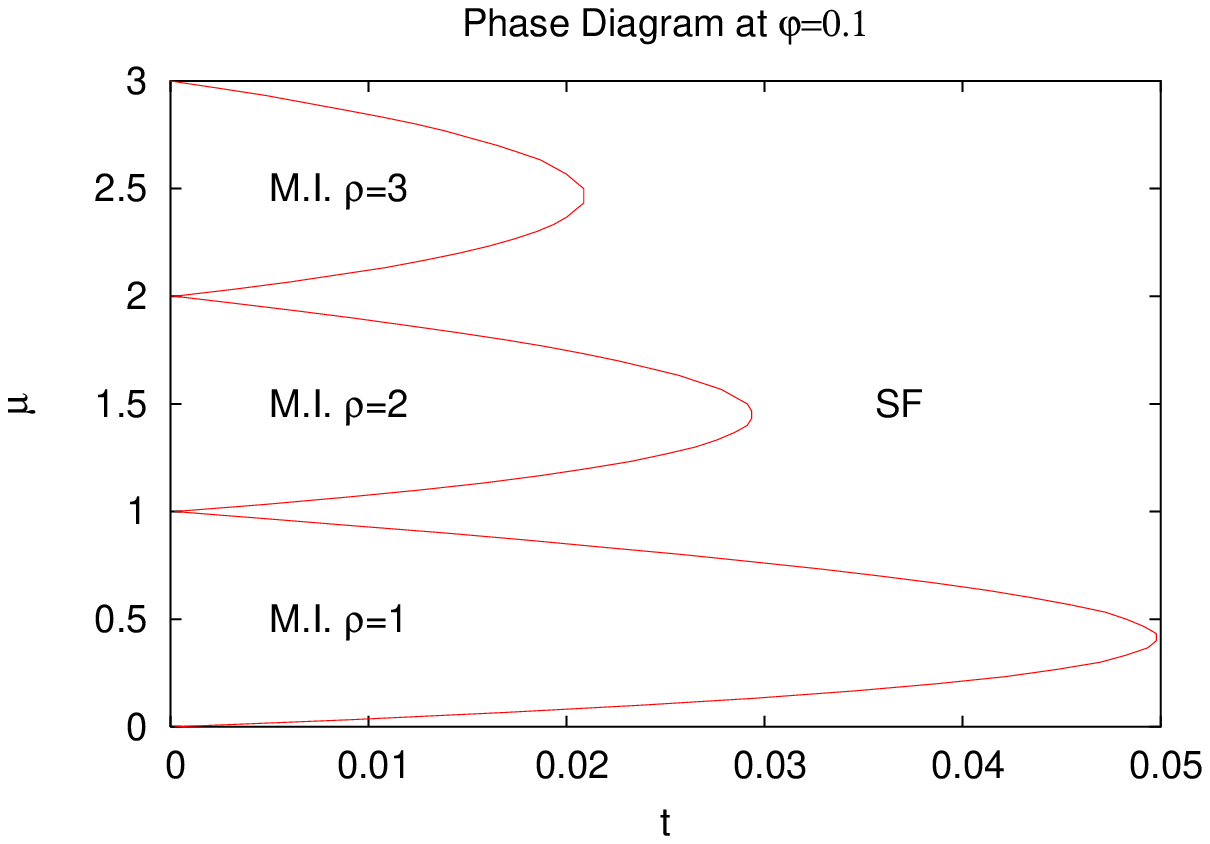}
\caption{(color online) Phase diagram of Bose atoms confined in
a 2D optical lattice for the magnetic flux $\varphi=0$ (left panel)
and 0.1 (right panel). In the figure the first three Mott lobes are
depicted, for on-site particle numbers 1, 2 and 3.}
\label{pd1}
\end{figure}

We denote by $\rho$ and $\rho_s$ the surface average of $\rho(nm)$
and $\rho_s(nm)$ respectively.
For a noninteracting system at zero temperature all of the bosons
are in the condensate, i.e.  in the lowest single particle state of
the lattice and we have $\rho=\rho_s$. When
the interaction increases (nonzero values of $U$ in Eq.\,(\ref{bhh}))
only a significant fraction of the bosons will condense in the same
single particle quantum state, and we have
$\rho_s<\rho$. The competition between the kinetic energy of the system
$t$ and interaction $U$ gives rise to interesting successive
transitions between a superfluid and a Mott insulator.

The mean-field solution of the nonmagnetic system and the
phase diagram are calculated by Sheshadri
{\it et al}.\cite{sheshadri93}. See also the Mott
insulator lobes in Fig.\,\ref{pd1} and \ref{pd3d00}.
For the magnetic Hamiltonian, perturbative techniques are used
by Niemeyer {\it et al}. \cite{niemeyer99}, where the MI lobes are
calculated for small values of the magnetic flux $\varphi=0,...,0.125$.

Introducing the magnetic field in the hopping term of the
Hamiltonian of Eq.\,(\ref{bhh}) breaks the temporal invariance of
the Hamiltonian and gives rise to persistent current flow of the
"charged bosons". For any bond connecting neighboring sites
$(i;k)=(n,m;n\pm 1,m)$ (or $(i;k)=(n,m;n,m\pm 1)$) of the lattice,
we define $t_{ik}=t$ for hopping along $x$ (or $t_{ik}=t e^{i 2
\pi n \varphi}$ for hopping along $y$). We calculate the local bond current
of the superfluid phase using the following formula:
\bea
\label{current} v_{ik}=\frac{1}{i\hbar}\left[t_{ik}a_i^\dagger
a_k -t_{ki}a_k^\dagger a_i\right]\, ,
\eea
by substituting $\langle a_i^\dagger a_k\rangle$ with $\langle \Psi_i
\Psi^*_k\rangle$.  Another parameter of
interest is the magnetic momentum. In the mean-field decoupling we
can define the single site magnetization by the following formula:
\bea\label{magnetization}
M_{nm}=n \bar{v}^x(nm)-m\bar{v}^y{(nm)}=-\frac{1}{\pi\hbar}\frac{d\Omega_{nm}}{d\phi}
\eea
where $\Omega_{nm}$ is the average value of the mean-field
Hamiltonian of Eq.\,(\ref{mf}) with respect to the ground state
Eq.\,(\ref{ssgs}) and represents the single site energy of the
Bose gas. The averaged site velocities $\bar{v}^{x,y}(nm)$ are equal to the average of the
bond currents of Eq.\,(\ref{current}) connecting the site $n,m$ to its neighbors along $x$ or $y$ accordingly.
The  magnetic momentum denoted with $M$ is equal to the surface
average of Eq.\,(\ref{magnetization}).

For zero magnetic field the 2D lattice has the translational
invariance along both axes and
the order parameter $\Psi_{nm}$ does not depend on the site index
$(nm)$.

In our case, for nonzero magnetic field in the chosen Landau gauge
the system preserves only the invariance along the
$y$ axis of the lattice. Therefore, the order parameter is chosen as
$\Psi_{nm}=\Psi_{n}$. In this case,
we calculate the mean-field solution for a given ratio of the magnetic
flux $\phi=p/q$. From the equation of the matrix elements of the
mean-field Eq.\,(\ref{men}), it can be noted that the periodicity of the
solution is $\Psi_{n}=\Psi_{n+q}$.
The same periodicity condition is verified by the
the density $\rho$ that also shows the translational invariance
along the $y$ axis: $\rho(nm)=\rho(n)$ and $\rho(n)=\rho(n+q)$.
To solve the mean-field equations we choose a finite
sequence of the lattice of dimension $q$
in $x$-direction and impose periodic boundary conditions.
In the Landau gauge, the system is periodic with lattice periodicity
in the $y$ direction, thus our calculations are carried out on a
$1 \times q$ supercell with periodic boundary conditions.
We measure the energy in units of $U$ (i.e. $U=1$).

\section{Results and Discussion}

\begin{figure}
\includegraphics[scale=0.7]{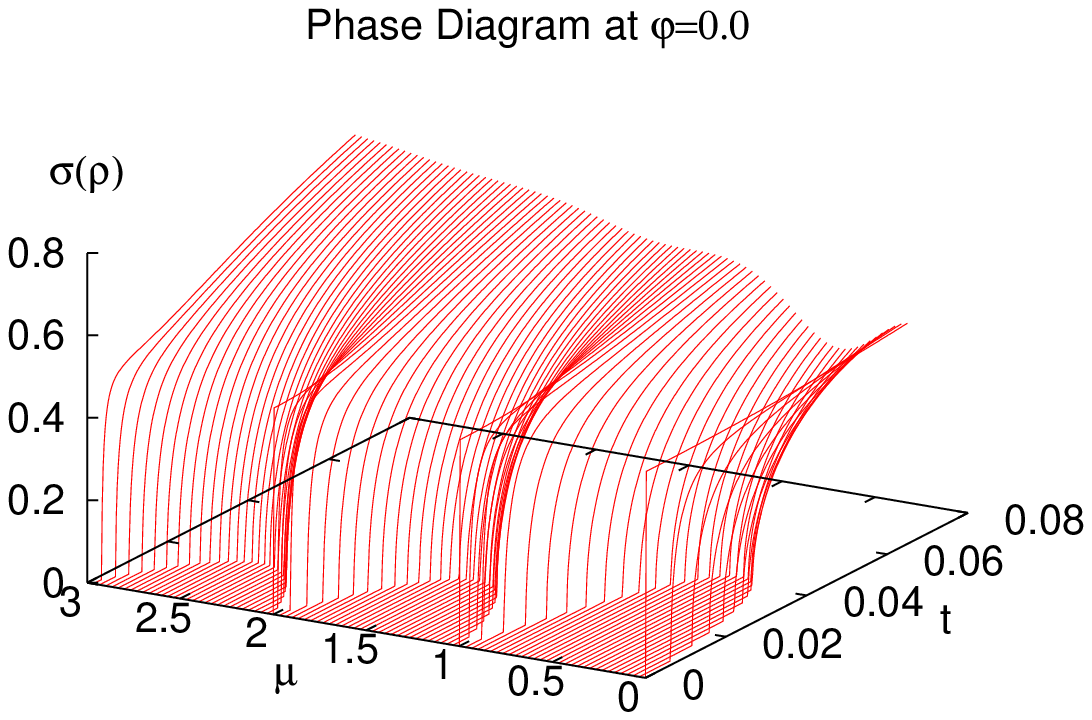}
\includegraphics[scale=0.7]{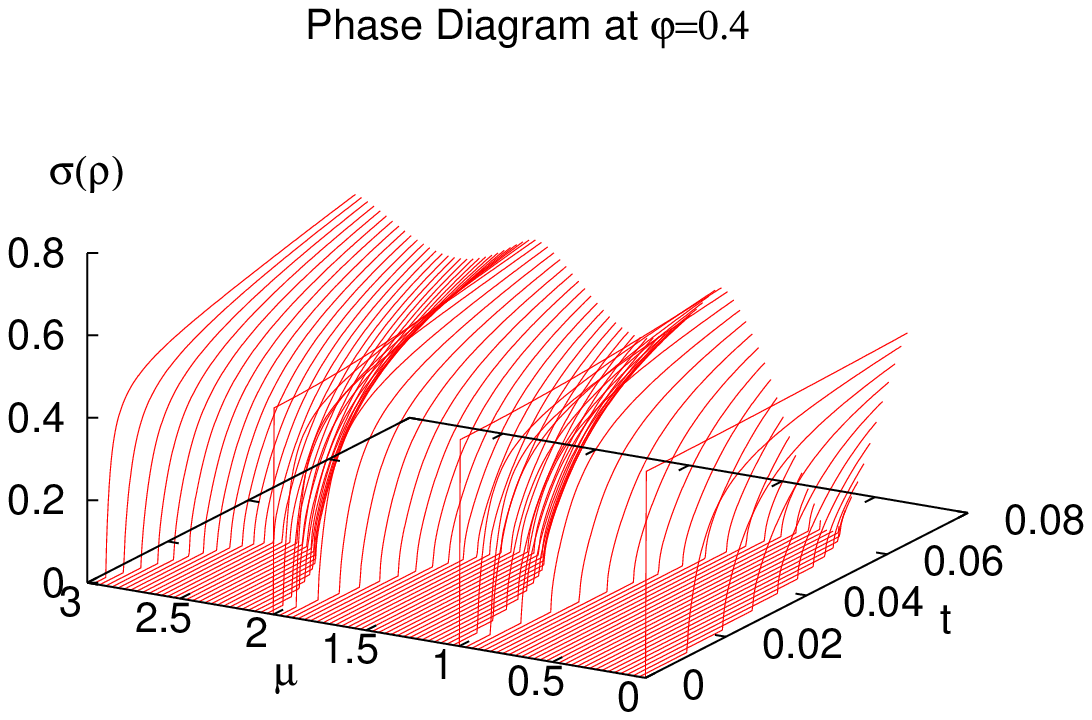}
\caption{(color online) Phase diagram for the magnetic flux $\varphi=0$ (right panel)
and 0.4 (left panel). On the $z$ axis
the variance $\sigma (\rho)$ of the Bose density
$\rho$ is plotted. For the MI phase the variance of the 2D density
is zero, and the Bose charge has no fluctuation.}

\label{pd3d00}

\end{figure}

Based on our calculations outlined in the preceding section,
we now present our numerical results. The two possible states of the
2D Bose system are selected as follows: in the MI phase, the on-site
occupation number has integer values and the variance of $\rho$ is
zero; in the SF phase
the on site occupation number has noninteger values and
the variance $\sigma (\rho)\ne 0$. The main features of our
results are illustrated in Figs.\,1-6.

\begin{figure}

\includegraphics[]{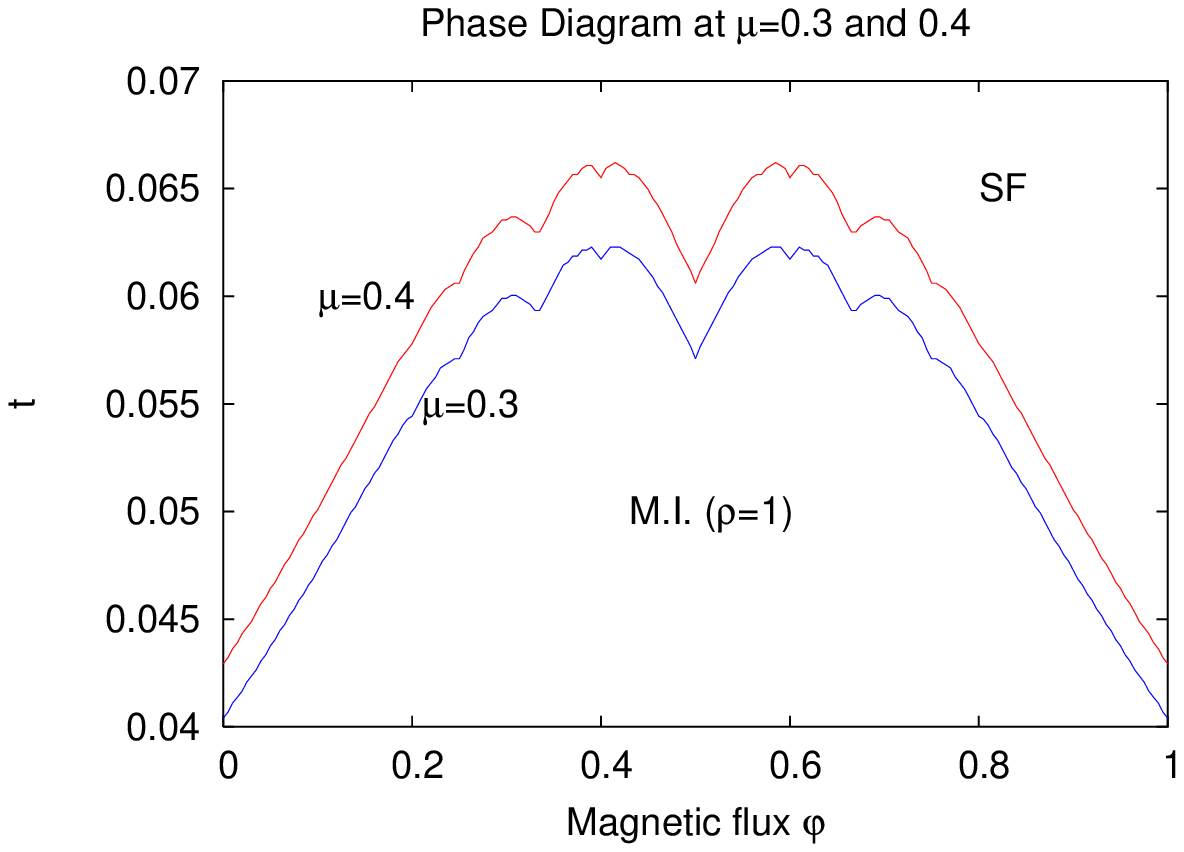}

\includegraphics[]{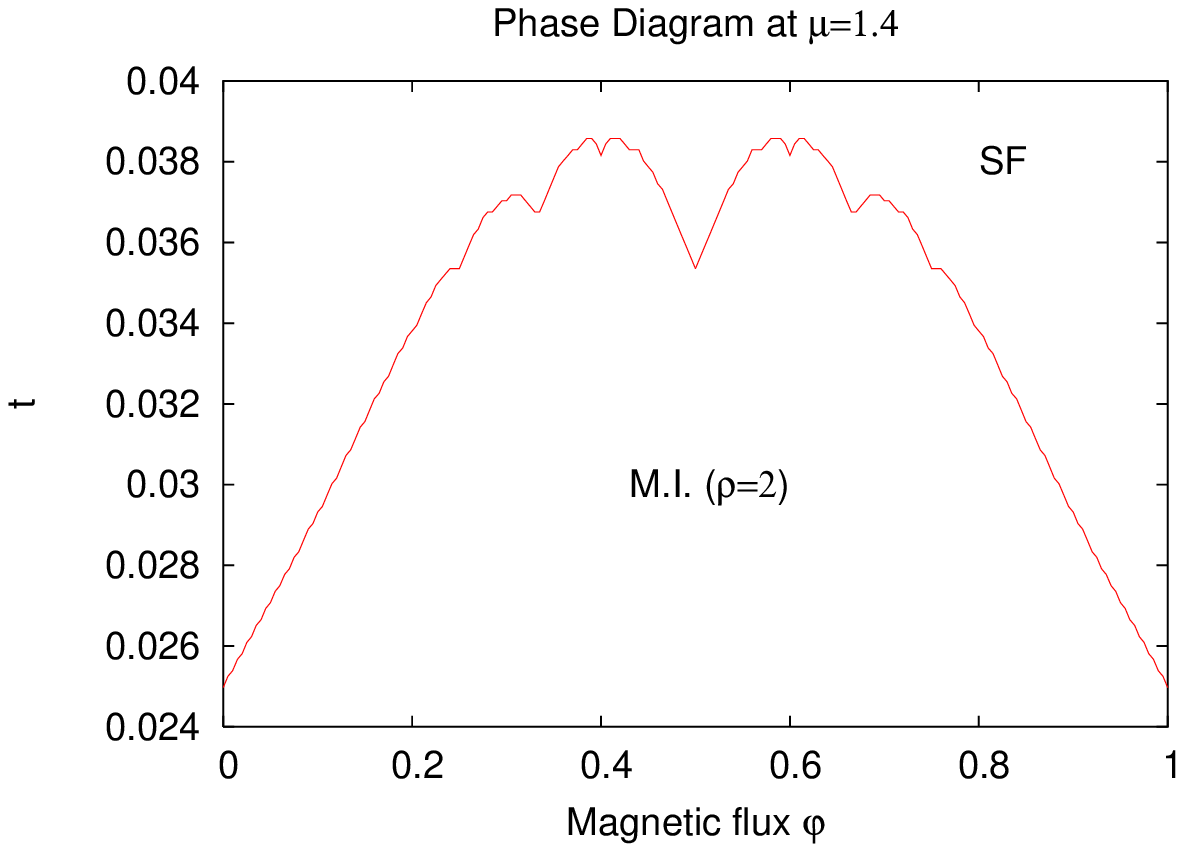}

\caption{(color online) Phase diagram of Bose atoms confined in
a 2D optical lattice in the $t-\varphi$
plane for different values of the chemical potential, corresponding
to the MI phase with $\rho=1$ (top) and $\rho=2$ (bottom).
The curves  correspond to the critical values $t_c$ of the SF-MI
transition. For $t>t_c$ the system is in the superfluid phase and
for $t<t_c$ in the MI phase. The phase diagram is periodic in the
magnetic flux with $\Delta\varphi=1$ and symmetric around
the flux value $\varphi=0.5$.}

\label{pdn1}

\end{figure}
\begin{figure}
\includegraphics[scale=0.7]{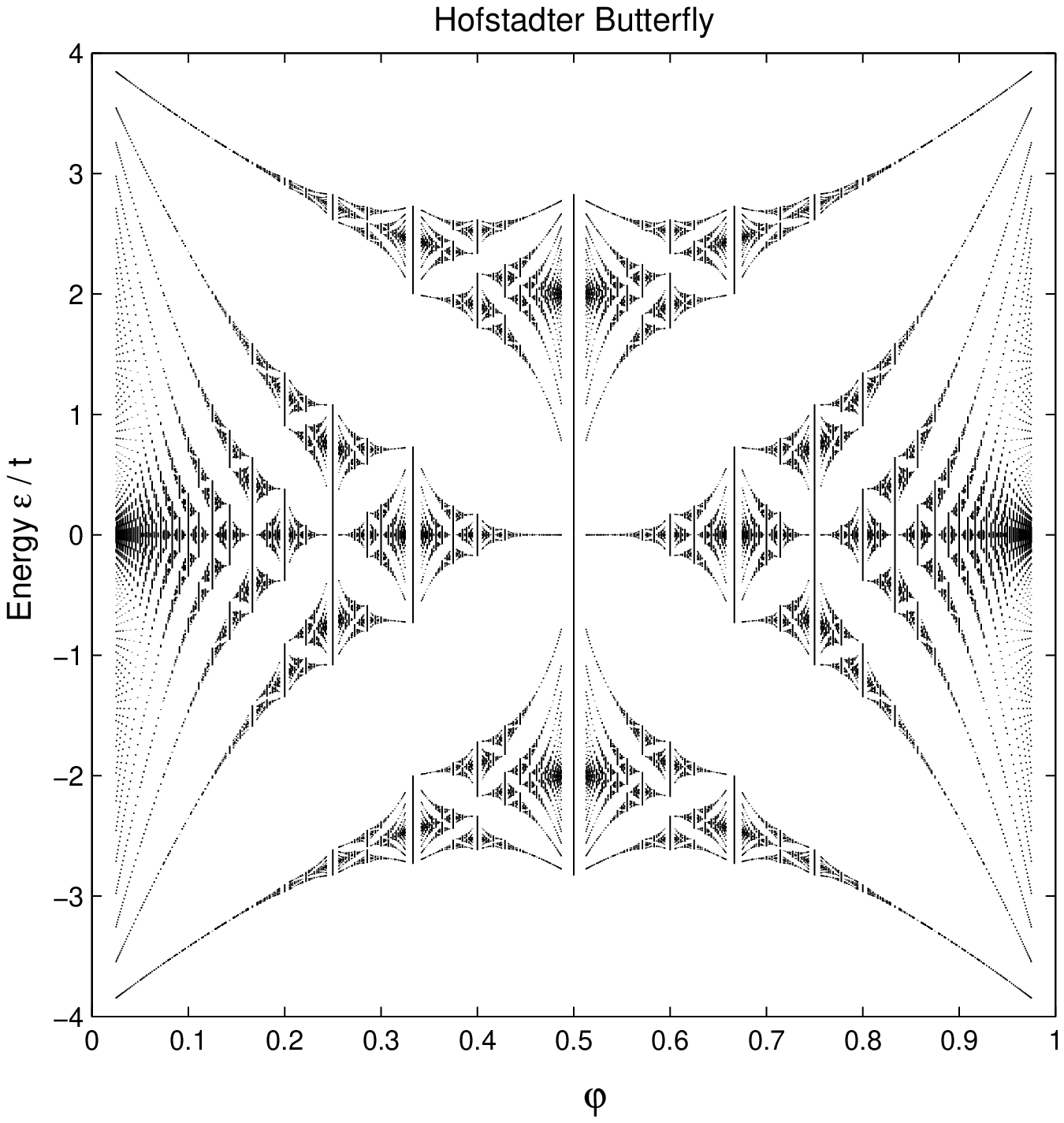}
\caption{ Single particle energy bands as a function of magnetic
flux per plaquette, the Hofstadter butterfly \cite{butter}. }

\label{butterfly}

\end{figure}

The phase diagram in the $\mu ,t$ plane is depicted in Fig.\ref{pd1}
which is calculated for $\varphi=0$ and $\varphi=0.1$.
For the Mott insulator state the site occupation number $\rho(nm)$
is equal to $\rho$ and has integer values (see Mott lobes in
Fig.\,\ref{pd1} for $\rho=1,\,2,\,3$). The variance of $\rho$
is equal to zero for the MI state
and has large fluctuations for the SF state (see the variance of
$\rho$ in Fig.\,\ref{pd3d00}).

The Bose-Hubbard model under a magnetic field has also been
considered by Niemeyer, Freericks and Monien \cite{niemeyer99},
using strong coupling expansion.
Strong coupling expansion utilizes a perturbative expansion in
$t/U$, and is valid within the Mott insulating regime for small
$t/U$ and small values of the flux $\varphi$.
The mean-field approach of this paper, on the other hand, is a
self-consistent but uncontrolled approximation, which
makes it possible to calculate physical quantities for both the
insulating and the superfluid regimes.
Although the two approximation methods have very different character,
they generally yield qualitatively similar results.
Very similar phase diagrams have been obtained by these two different
approaches for the pure Bose-Hubbard model
(\cite{sheshadri93} and \cite{SCEpure}),
and the  superlattice Bose-Hubbard model
(\cite{MFsuperlattice} and \cite{SCEsuperlattice}).
Indeed, we find that our mean-field treatment produces a qualitatively
similar result to the strong coupling expansion for the
Bose-Hubbard model under a magnetic field. However, our method
allows us to extend the calculation to the full range of magnetic
flux $0<\varphi<1$, and calculate the superfluid density, and
magnetization in the superfluid phase.

We note that the transition point at zero magnetic flux is located at
$t_c=0.043$ for $\mu=0.5$ (c.f. Fig.\,\ref{pd1}).
This corresponds to $U/4t =5.8$ which is equal to the transition
point determined by van Oosten {\it et al}. \cite{oosten03}.


The magnetic flux breaks the temporal invariance of the system and
destroys the coherence of the wave function in a 2D fermion system
after a "flight" time proportional\cite{bergmann84} to $1/B$.
For a Bose system, the magnetic flux can have the same effect over
the SF coherent wave function, destroying the stability of the
superfluid solution in the proximity of the transition point and
leading to the SF-MI transition when $\varphi$ is bigger than a
critical value $\varphi_c$.
This is consistent with the increased area of the Mott lobes when
the magnetic field is present
as shown in Figs.\,\ref{pd1} and \ref{pd3d00}.
For all values of the magnetic flux we depicted the phase diagram
in the $t,\varphi$ plane for the different values of the chemical
potential in Fig.\,\ref{pdn1}.

We note the above MI-SF transition for a constant $t$ value,
increasing the magnetic flux $\varphi$ from zero to
the critical transition point $\varphi_c$.
For small $\varphi$ the optical lattice approximates a continuous
system of charged bosons under the magnetic field and the
superfluid-insulator transition is similar to the disappearance of
the superconductivity when the external magnetic flux in a
superconductor exceeding a critical value\cite{superconductor}.

\begin{figure}

\includegraphics[]{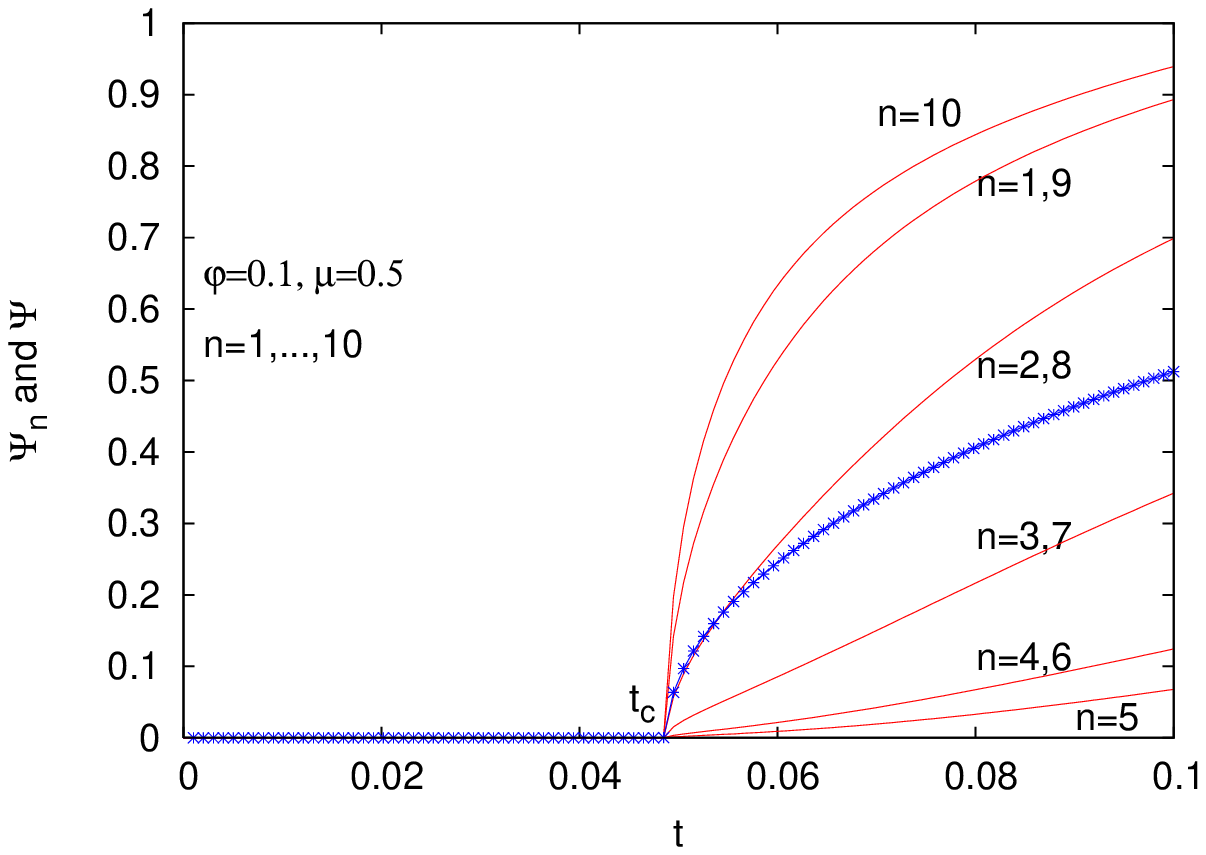}
\includegraphics[]{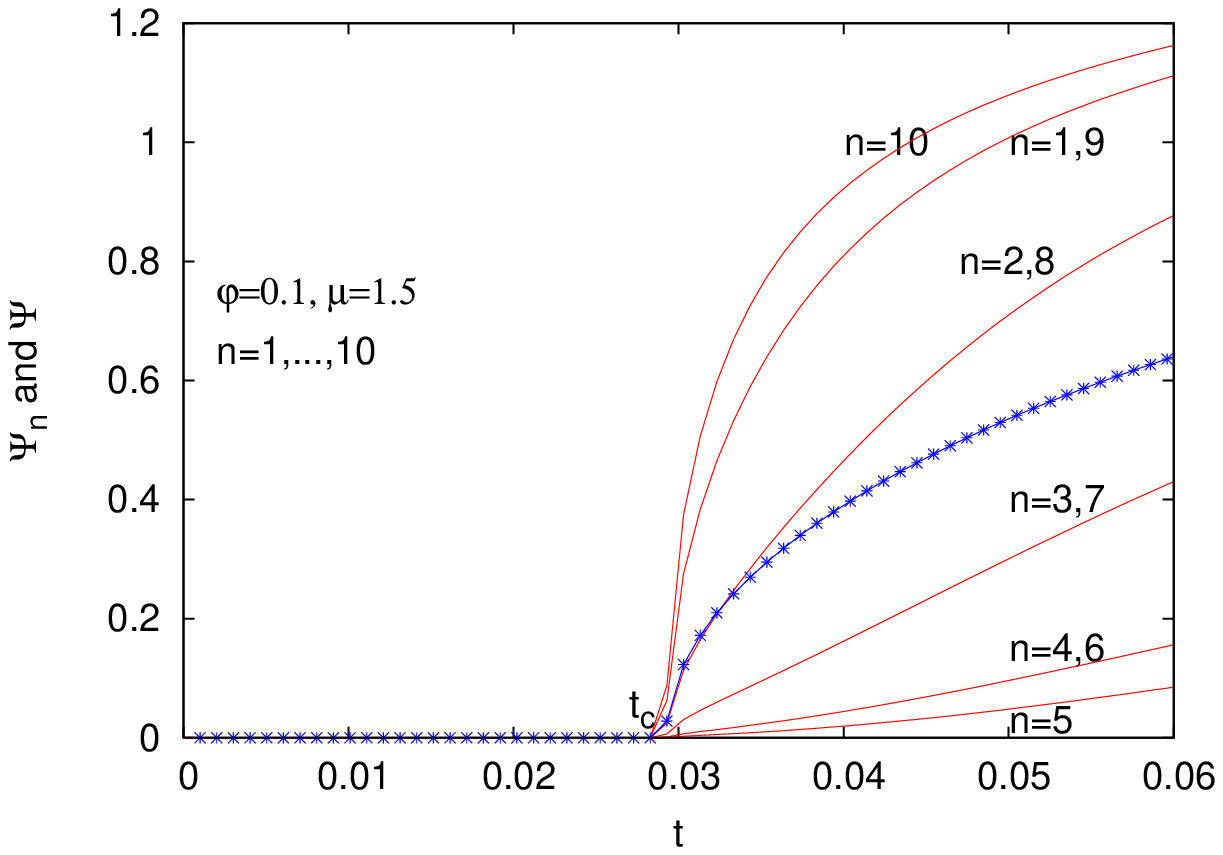}
\caption{(color online) SF order parameter $\Psi_n$ for different lattice sites at
MI-SF transition. The chemical potential is $\mu=0.5$ corresponding
to MI with $\rho=1$ and $\mu=1.5$ corresponding to MI with $\rho=2$.
The blue line with crosses represents
the surface average value $\Psi$. For $t>t_c$ the second derivative
$d^2\Psi/dt^2$ changes sign for the points $n=4,\,5$.
The magnetic flux is $\varphi=0.1$ and the
superfluid order parameter has the periodicity $\Psi_n=\Psi_{n+10}$.
For the SF phase (at $t>t_c$) the lattice sites $integer*10+4$ and
$integer*10+5$ exhibit very
low values of the SF order parameter and low variance of $\rho$.}

\label{detail1}

\end{figure}

\begin{figure}
\includegraphics[]{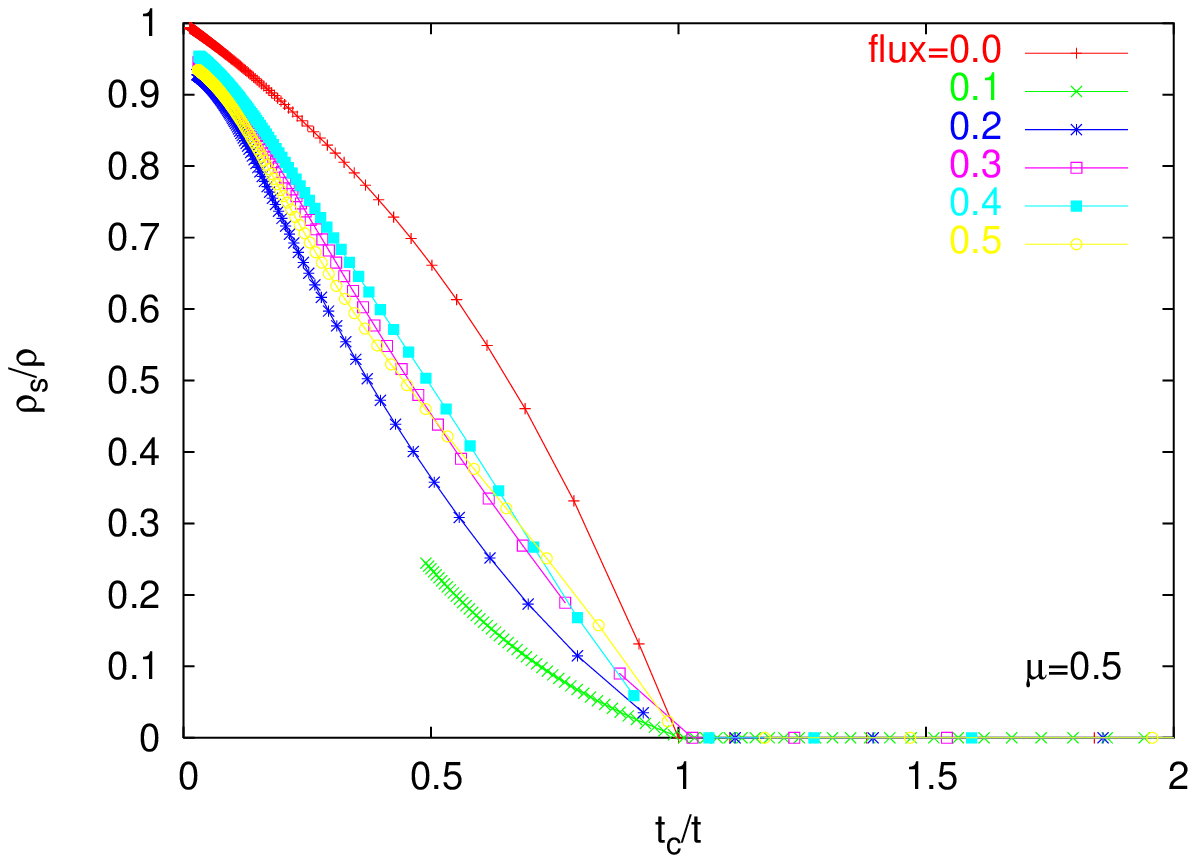}
\caption{(color online) $\rho_s / \rho$ as a function of $t_c/t$
for different values of the magnetic flux. For $\varphi=0$ our
curve is similar to Fig.\,1 of Ref.\,\cite{sheshadri93}. The
change of slope at $t\simeq > t_c$ for $\varphi \neq 0$ is related
to the surface oscillations of the charged bosons. When the
magnetic field is present, the superfluid phase exhibits  surface
region where the quantum fluctuations of SF density and the order
parameter are small (compared to nearby regions; see
Fig.\,\ref{detail1}) meaning that the local phase is closer to an
insulator. It exhibit a lower ratio $\rho_s / \rho$ compared with
the case $\varphi=0$.
}
\label{comp}
\end{figure}

However, for the Bose-Hubbard Hamiltonian the discreteness of the
lattice brings about other interesting features. The noninteracting
spectrum\cite{butter} is periodic in $\varphi$ with $\Delta\varphi=1$
and symmetric around the value $\varphi=0.5$. We recover the same
feature of the phase diagram in Fig.\,\ref{pdn1}.

One of the most important results of our calculation is displayed
in  Fig.\,\ref{pdn1}. The oscillations of the critical hopping
strength with changing magnetic field follow the oscillations of
the bandwidth of the Hofstadter butterfly\cite{butter}
(See Fig.\,\ref{butterfly}).
For instance, increasing $\varphi$ the bandwidth of the Hofstadter
butterfly system \cite{butter}
is shrunk, meaning that the superfluid order appears at a smaller
value of $U$ (or higher value of $t$). For $\varphi\to 0.5$ the
bandwidth of the Hofstadter butterfly
is increased again and the critical value $t_c$ of the transition
is increased. This suggests that in the phase diagram of
Fig.\,\ref{pdn1} (calculated for $U=1$),
the critical point $(U/t)_c$ of the SF-MI transition
is proportional to the bandwidth of the Hofstadter
butterfly\cite{butter}.

\begin{figure}

\includegraphics[scale=1]{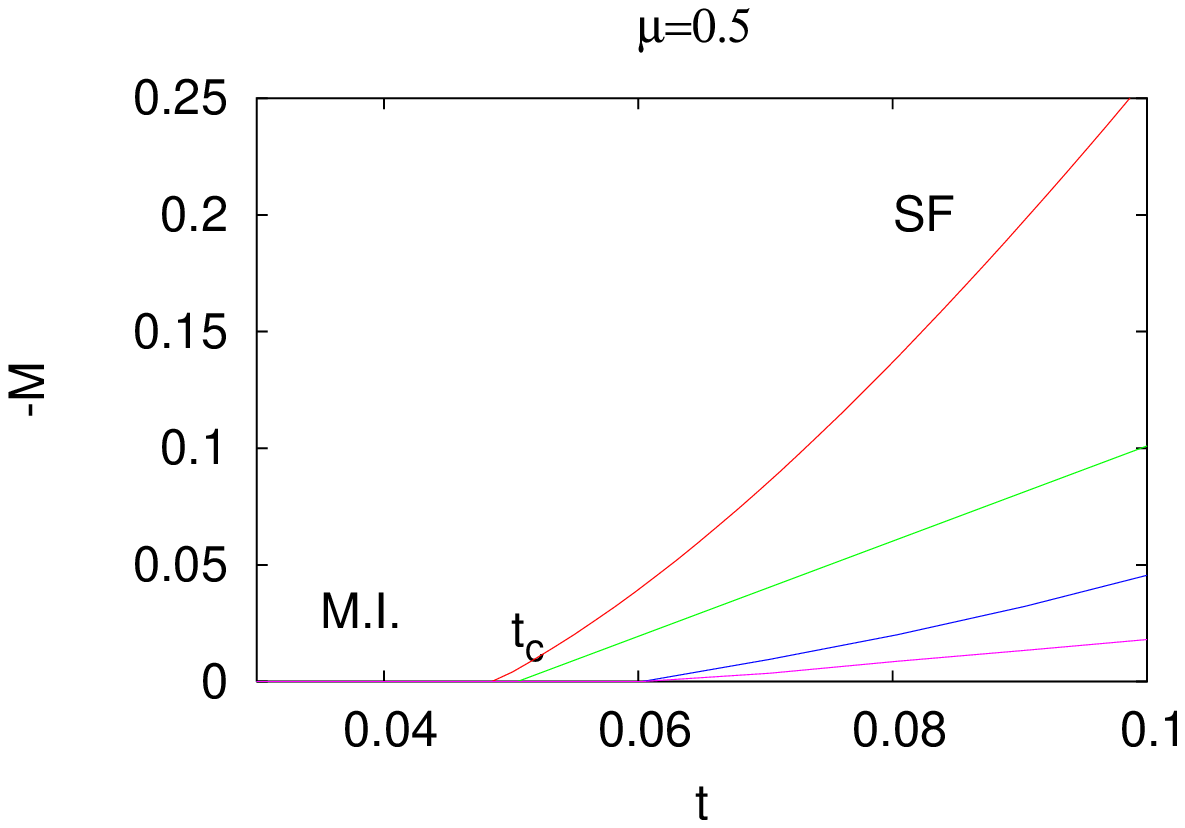}

\caption{(color online) Magnetization as a function of $t$ for
$\varphi=$ 0.1, 0.2, 0.3, 0.4 (from top to bottom).
In the MI phase, for $t<t_c$, the magnetization is zero. Our numerical approach
is more reliable at lower values of $t$, and not too small $\varphi$,
where our computational supercell size is small and the number fluctuations at each lattice site
are limited. MI and superfluid phases and $t_c$ are identified for  $\varphi=0.1$ on the figure. The value of
$t_c$ for other $\varphi$ can be identified as the point where magnetization becomes non-zero on the corresponding curve.
 }
\label{mgn}
\end{figure}


The loss of translational invariance of the charged bosons under a
magnetic field is correlated with the appearance of the surface
oscillations of the particle density in the superfluid phase.
Although one expects a spontaneous breaking of the translational
invariance in the superfluid phase, such as vortex states, in our
mean-field approach this symmetry is explicitly broken by our
gauge choice. Physically measurable quantities should be independent
of the choice of the gauge, however our mean-field treatment yields
spatially dependent parameters such as $|\Psi_n|^2$, which
depend on the gauge choice. Still, it is not unreasonable to expect
that spatially averaged quantities such as $\rho_s$ and magnetization
to be correctly captured by our approach. For certain values of the
magnetic field we verified this expectation by using the symmetric
gauge and corresponding square supercell.
Also, as the insulator side of the
transition is spatially uniform, explicit determination of the
gauge should not strongly affect the MI-superfluid phase boundaries.
In Fig.\,\ref{detail1} we show the variance of the on-site SF order
parameter $\Psi_n$ when the hopping parameter $t$ is increased,
for different points of the lattice $n$. Even all the SF order
parameters $\Psi_n$ exhibit the disappearance of the insulator
order at the same critical value, $t=t_c$, the rate of increase of
$\Psi_n$ for $t>t_c$ is different for different points of the
lattice. There are points in the lattice that show a very low rate
of increase of $\Psi_n$. The figure clearly shows the surface
oscillations of the 'charge' density in the optical lattice. It
means that, for $t>t_c$, there are regions in the lattice with
small SF order parameter and small density fluctuations. It is
possible to interpret these as the superfluid density oscillations
due to the presence of vortices. However, close to the Mott
transition where these regions are most pronounced, correlations
between such regions may develop causing a phase transition which
would not be captured by our mean-field treatment.

The lower rate
of increase of the superfluid order parameter vs. $t$, for
$t>t_c$, is also suggested in the scaled curves in
Fig.\,\ref{comp} that show the ratio $\rho/ \rho_s$ versus
$t_c/t$. For $t_c/t\simeq<1$, the SF phase is strongly affected by
the magnetic field presence, and we note the change of the curve
slope in Fig.\,\ref{comp} for $\varphi\ne 0$. For $t_c/t\to 0$,
the magnetic field has no effect, and all Bose particles tend to
condense ($\rho/ \rho_s\to 1$ for all $\varphi$ values in
Fig.\,\ref{comp}). We also notice that even a small magnetic field
$\varphi=0.1$ affects the superfluid density rather strongly.
For such small magnetic fields our approach is less reliable,
as the cell size used in our calculations
is inversely proportional to the flux $\varphi$.
While similar discontinuous effects at zero
field have been discussed within the context of Josephson junction
arrays \cite{Josephson}, we can not clarify
the behavior near zero field due to the limitations of our numerical
mean-field approach.


Figure \ref{mgn} shows the magnetization
as a function of the hopping parameter $t$ for different
values of the magnetic flux $\varphi=0.1,\,0.2,\,0.3$ and $=0.4$
at the chemical potential $\mu=0.5$. We note that our
approach is less reliable for the case of $\varphi=0.1$, as the
number of particles per site increase most quickly in this case.
For $t<t_c$, in the MI phase, the magnetization is equal to zero
and the particle motion is frozen.
For $t>t_c$, in the SF phase,
the curves exhibit negative values of magnetization,
meaning the existence of persistent current flow.
We note that the sign of the magnetization is related to the slope
of the energy curve vs. magnetic flux in Hofstadter
butterfly \cite{mine}.

The change of slope of the eigenstates as a function of $\varphi$ in the
Hofstadter butterfly gives rise to changes in the sign of the
magnetization near the special values of the magnetic flux
(for instance $\varphi=1/q$; see Ref.\,\cite{butter}).
We do not expect this fine effect to be observable in the
low $\varphi$ limit. In this case, the small fluctuations of the
$\varphi$ for the real system gives rise to the smeared graph of
the spectrum \cite{butter}.

For $\varphi\in[0.45,0.5]$ the magnetization can have an opposite
sign (compared with the values for $\varphi\in [0,0.45]$) as the
slope of the Hofstadter spectrum vs. $\varphi$ clearly changes
(see \cite{butter} or Fig.\,\ref{pdn1}).

\section{Summary}

We calculated the mean-field phase diagram of the 2D Bose-Hubbard
Hamiltonian under a perpendicular magnetic field. The Mott
insulator-superfluid transition is strongly affected, and the
features around the transition point resemble the interesting
property of the Hofstadter butterfly. The energy spectrum for
noninteracting case exhibits periodicity with $\Delta\varphi=1$,
symmetry around the value $\varphi=integer/2$ and striking
oscillations\cite{butter} that lead to similar features of the
MI-SF transition when the magnetic field is varied (see the phase
diagram in $t,\varphi$ plane in Fig.\,\ref{pdn1}).

In the superfluid phase, at zero magnetic field the system has
time translational invariance and the net local current of the
Bose particles is zero because the reversed Bose paths are equally
probable. Even for a small value of $\varphi$ the time invariance
is suddenly broken and the reversal paths of the coherent Bose
atoms are not equally probable anymore (in fact only one of the
two reversal paths is permitted; the other one corresponds to
change of the sign of the magnetic flux). The persistent currents
of the Bose particles appear, leading to nonzero value of the
orbital magnetization. We note also the surface charge
oscillations, that are not present in the SF phase with
translational invariance when the magnetic field is set to zero.

The surface charge oscillations are not present for MI phase. At
low values of $\varphi$ the SF phase exhibit negative
susceptibility when the magnetic field is applied and at a
critical value the system is driven into the insulating phase.
This is the continuous limit of the model and the behavior of the
Bose system shows features resembling to that of the Meissner
effect in superconductors. For higher kinetic energy (large values
of $t/U$) this scenario is not valid anymore, the magnetic flux
loses its effect over the quantum state of the Bose system and the
system preserves its SF characteristics, albeit with spatial
oscillations of its order paremeter. We hope our work stimulates
further experimental and theoretical interest in this model.

\acknowledgments{M.\,N. acknowledges the support from TUBITAK
and CERES and thanks the Department of Physics, Bilkent University
for hospitality during the time part of this work was performed.
M.{\"O}.\,O is supported by TUBA-GEBIP grant and TUBITAK-Kariyer
grant No. 104T165. B.\,T. gratefully
acknowledges partial support from TUBITAK and TUBA.}

\end{document}